\documentclass[preprint,authoryear,12pt]{elsarticle}
 \usepackage{graphicx}
\usepackage{amssymb}
\journal{Planetary and Space Science}
\begin{document}

\begin{frontmatter}
\title{Remote Sensing of Chiral Signatures on Mars}

\author[wbs]{William Sparks}
\address[wbs]{Space Telescope Science Institute, 3700 San Martin Drive, Baltimore, MD 21218, USA. (sparks@stsci.edu)}
\author[jhh]{James H. Hough}
\address[jhh]{University of Hertfordshire, College Lane, Hatfield, AL10 9AB, UK. (j.h.hough@herts.ac.uk)}
\author[tag]{Thomas A. Germer}
\address[tag]{National Institute of Standards and Technology,  100 Bureau Drive, Gaithersburg, MD 20899, USA. (thomas.germer@nist.gov)}
\author[fr]{Frank Robb}
\address[fr]{Institute of Marine and Environmental Technology, University of Maryland School of Medicine, 701 East Pratt Street, Baltimore, MD 21202, USA. (FRobb@som.umaryland.edu)}
\author[lk]{Ludmilla Kolokolova}
\address[lk]{Department of Astronomy, Universiy of Maryland, College Park, MD 20742, USA. (ludmilla@astro.umd.edu)}

\begin{abstract}
We describe circular polarization as a remote sensing diagnostic of chiral signatures which may be applied to Mars. The remarkable phenomenon of homochirality provides a unique biosignature which can be amenable to remote sensing through circular polarization spectroscopy. The natural tendency of microbes to congregate in close knit communities would be beneficial for such a survey. Observations of selected areas of the Mars surface could reveal chiral signatures and hence explore the possibility of extant or preserved biological material. We describe a new instrumental technique that may enable observations of this form.
\end{abstract}


\end{frontmatter}


\section{Introduction}
\label{introduction}

Photosynthesis is a remarkable life strategy offering enormous evolutionary advantages.  It is reasonable to suppose that if it gets chance to gain an evolutionary foothold, it will do so on worlds other than the Earth, including Mars. On Earth, microbial photosynthesis arose early, and quickly became dominant to the extent that the global atmospheric composition of Earth was modified. In the ancient past of Mars, it is widely believed that conditions were not unlike those on early Earth, with abundant water and energy from a variety of sources, including the Sun. If life also originated on early Mars, and followed a similar trajectory to life on Earth in the beginning, then microbial photosynthesis may very well have evolved in the ancient water basins of Mars. Subsequently, the Martian and terrestrial physical characteristics diverged, with Mars becoming, cold, arid and bathed in ultraviolet radiation, an environment hostile to life as we know it on Earth. Nevertheless,  life has a remarkable capacity for adaptation, and throughout Martian history, free energy has been available in abundance from the Sun. While terrestrial organisms would not survive on the Martian surface, there are situations on Earth that indicate mechanisms  exist which would protect microbial organisms on Mars should they have had the opportunity to adapt. These include protective pigmentation against ultraviolet radiation \citep{Cockell_1999, Cockell2003}, inclusion of antifreeze in the cellular fluids \citep{Carpenter1992},
a resistance to the stresses introduced by a high saline content \citep{Litchfield1998, Landis2001}, and endolithic residence beneath a thin protective layer of translucent rock \citep{Norris2006, Wierzchos2006}. Life is also skilled at taking advantage of localized niches of habitability and nutrients which may be present
\citep{Davey2000, HallStoodley2004}. Hence, we suggest it is possible that extant or preserved photosynthetic life may be evident at or near the surface of Mars.

A high quality biosignature arises uniquely from biological processes. If a biosignature can additionally be detected with remote sensing, then it can be useful to survey extensive surface areas to identify interesting regions for in-depth study, either remotely or with landers. This is a substantial advantage relative to in situ measurements where only a tiny fraction of the planet may be studied. The remarkable phenomenon of biological homochirality, whereby the chiral building blocks of life use practically exclusively L-amino acids and D-sugars, may be such a biosignature. Furthermore, the optical activity of typical biological molecules coupled with their homochirality can yield a measurable polarization signal. Hence, homochirality, a powerful biosignature, can be amenable to remote sensing through circular polarization spectroscopy, a standard tool in analytical chemistry, e.g. studying protein structures  \citep{Purdie1994, Kelly2000}. Spectropolarimetry is a familiar tool to astronomers, and hence relatively conventional instrumentation can be used to survey the surface of Mars in this manner \citep{Sparks2005}. For a future space-based mission such as an orbiter, more innovative approaches are likely to be required, such as the robust, full Stokes static optics method of  \citep{Sparks2012ao}.

Since photosynthesis is an inherently surface or near-surface phenomenon, using visible light and requiring a  transparent atmosphere, and is a strong polarization-sensitive physical interaction between light and the organism, it is optimally observable if present. Circular polarization spectroscopy therefore offers a feasible  remote sensing diagnostic technique to search for the presence of (chiral) photosynthetic microorganisms. Observations of potentially vast areas of the Mars surface could be undertaken, revealing chiral signatures if present, and hence exploring the possibility of extant or preserved biological material.

Here, in \S~2, we describe laboratory measurements  to support this concept, \S~3,  a brief review of previous efforts to measure the polarization of the Mars surface, and 
\S~4, a concept for a new approach to seeking chiral signatures on the surface of Mars.

\section{Laboratory Work: Remotely Sensing Life's Signatures}
\label{labwork}

Photosynthesis operates by means of a strong electronic transition in  chiral photosensitive pigments. We carried out laboratory tests to determine whether this resulted in a measurable circular polarization signature in the spectrum of light scattered from photosynthetic micro-organisms and vegetation. Extensive measurements have been carried out in two laboratories, at the National Institute of Standards and Technology (NIST, Maryland) and the University of Hertfordshire (UK). This work was published in \cite{Sparks2009pnas, Sparks2009jqsrt} and \cite{Martin2010} who presented spectropolarimetry of a variety of photosynthetic microorganisms, leaves and control mineral samples.

Theoretically, and empirically, circular polarization can arise whenever mirror symmetry is broken. This can occur in a variety of non-biological situations, including non-thermal synchrotron emission, multiple scattering,  scattering of polarized light from aligned grains, transmission of radiation through a medium of grains whose alignment twists along the line of sight, and cyclotron radiation. Multiple reflection and geometrical phase effects can introduce circular polarization; however, they
produce a smooth intensity distribution recognizably related to the geometry of the scattering. Furthermore, scattering processes lack the spectral correlation of polarization with absorption found in biologically produced circular polarization \citep{deg1992, barron2008, Sparks2009pnas}.

\citet{Sazonov1972} considered the causes of circular polarization in the case of the light scattered by a surface and showed, based on a fundamental consideration of the P-T invariance of electromagnetic interaction that non-zero circular polarization can be produced in the case when ``the reflecting medium is non-gyrotropic.'' This can result from the influence of the magnetic field, presence of crystal with no center of symmetry or homochiral molecules. He also noticed that an asymmetric medium (e.g. a surface with some topographic features at a non-zenith illumination) can become non-gyrotropic due to multiple scattering. However, in this case the scattering material should be either absorbing or the reflection should happen at an angle that exceeds the angle of the total internal reflection. The last opportunity looks very unlikely in the Martian conditions. The other case, multiple scattering by absorbing materials, was checked in laboratory measurements. \citep{deg1992} that showed that circular polarization stays zero for the light scattering by oblique layers (to mimic asymmetric illumination) of transparent particles (e.g. quartz) but becomes non-zero for absorbing materials and especially in the case of highly absorbing metals.  \citet{deg1992} could reproduce the measured curves theoretically using a model of multiple Fresnel reflections.

The laboratory work carried out at NIST and the University of Hertfordshire found significant circular polarization in the electronic absorption bands of both primary and antenna photosynthetic pigments (Fig.~1, adapted from \citet{Sparks2009pnas}). In addition to the biological samples, however, we acquired a suite of measurements of abiotic controls. Fig.~2 consolidates the control measurements quantitatively, and displays all three of these important tests using the same scale and conventions. 
Sulphur and iron oxide were chosen as both have strong spectral features reminiscent of the red-edge of chlorophyll which causes substantial structure in the
polarization signal around the edge. It is clear from Fig.~2, however, that despite the strength of the spectral edge, neither mineral shows any polarization structure correlated with the spectral features. Reassuringly, the Mars regolith simulant JSC-1 shows structure in neither the spectrum nor the polarization spectrum, very different to the microbial samples. As well as the degrees of circular polarization, which can be produced by multiple scattering in combination with geometric effects, being very low for the control samples, the scattered spectrum did not contain any of the key spectral features associated with the microorganisms.

\section{Telescopic Observations}
\label{ground}

To probe the surface of Mars for chiral signatures is clearly speculative yet, with current advances in the field, technically feasible. Resolving a chiral signature that is localized on the Martian surface is challenging and the prospects can be summarized in the following. High-sensitivity polarimetry is possible from ground-based telescopes \citep{Hough2006}, making such observations relatively inexpensive. Wide spectral coverage (350~nm to 1000~nm) is desirable but this is readily achievable with most observatory spectrometers.

Though the biological signals are several orders of magnitude larger than the polarization limits in the control samples, they are nevertheless quite small in an absolute sense, in the range $10^{-2}$--$10^{-4}$ for polarization degree. Polarimetric measurements, however, are capable of very high sensitivity.  For example \cite{Hough2006} describe an astronomical  polarimeter (PlanetPol) that achieves sensitivities for fractional linear polarization of better than 1~part per million. Such a polarimeter could equally well be configured to measure the circular polarization rather than linear. If the undiluted circular polarization strength is $\approx 0.1$\%\ as found for terrestrial microbial communities, and the polarimeter can measure to a polarization degree of $\sim 10^{-6}$, then it follows that for a good detection, the dilution factor that can be tolerated is $\approx 0.01$. That is, if 1\%\ of the target area is covered by a microbial community yielding a polarization signal of $10^{-3}$, then the net signal is a polarization degree of $10^{-5}$, detectable with a current-technology precision polarimeter.

High spatial resolution, therefore, is most important, so that intrinsically interesting sites can be observed without reducing the potential chiral signatures through a large amount of dilution due to light scattered from surrounding regions. An optical interferometer, such as the VLTI which has a goal of 0.002 arcsec, will achieve spatial resolutions of $\sim 500$~m.  However, such interferometers are used at near-infrared wavelengths and polarimetry would be technically quite challenging with such an interferometer.  The best that can be achieved with a single dish ground-based telescope, using adaptive optics, is (optimistically) $\sim 0.1$~arcsec, giving a spatial resolution of $\sim 25$~km during a favorable opposition, and  comparable to the size, e.g., of saline surface deposits \citep{Osterloo2008}.

Thus,  circular polarization signatures shown in Fig.~1 are observable with ground-based astronomical telescopes provided (\romannumeral1) the region of high chirality occupies of order 1\%, weighted by reflectance, of a region 25~km in extent, typical of saline deposits on Mars and (\romannumeral2) there are sufficient photons available to make the measurement in a small enough time period. This is  the case for large ground-based telescopes such as the VLT, and may be feasible with smaller telescopes by sacrificing some resolution in time and spatially. 

Early large aperture photopolarimetry circular polarization observations of Mars, Jupiter, and other planets have been made. The highest known integrated circular polarization detection in the Solar System is from Mercury which has a fractional polarization of $10^{-4}$ \citep{Kemp1971}, attributed to the handedness of quartz.
Other abiotic processes are discussed therein to explain the small levels of circular polarization found for the other planets. The Mars observations of \cite{Kemp1971}, taken far from opposition revealed a  degree of circular polarization of 2--$8\times 10^{-5}$ when the data were integrated over both hemispheres of the planet.
\cite{Wolstencroft1972} showed a globally integrated wavelength dependence of circular polarization for Mars revealing integrated polarization levels typically $2\times 10^{-5}$.

We obtained high-quality imaging circular polarimetry using the 8.2-m aperture European Southern Observatory Very Large Telescope (ESO VLT) of a portion of the Mars surface during the favorable opposition of 2003 to seek evidence of anomalous optical activity that might indicate large scale surface chiral phenomena. In these early observations the goal was to map as much of the Martian surface as possible rather than targeting specific locations. Mars was close to a favorable opposition at the time of the observations and the angular resolution of $\sim 0.8$~arcsec, achieved without any adaptive optics, corresponded to 200~km spatial resolution. Using two narrow
band filters, we did not find any areas of circular polarization at that scale \citep{Sparks2005}. 

Specifically, we used two widely separated narrow-band filters, 378 nm and 953 nm, covering 43\%\ of the martian surface, 15\%\ of it in-depth. With polarization noise levels $<0.1$\%\ ($4.3\sigma$ upper limits 0.2--0.3\%) and spatial resolution 210~km, we did not find any regions of circular polarization. When data were averaged over the observed face of the planet, we did see a small non-zero circular polarization 0.02\%, which may be due to effects associated with the opposition configuration. 
We wish to emphasize that these observations covered only a tiny fraction of parameter space. Only 14.6\%\ of the surface was covered, and only 1.7\%\ of the accessible spectral window yielding a total coverage of only 0.25\%\ of the available observational parameter space. Coupled to the low spatial resolution of these ground based observations, which would only be sensitive to signatures on a scale of $\sim 200$~km or larger, the study cannot be used to conclude there are no chiral surface signatures on Mars. Indeed, we are greatly encouraged by the absence of polarization features which is empirical evidence that ``false positives''  from abiotic mineralogical
or scattering processes are not prevalent in this method.

\section{A Concept for Future Exploration}
\label{future}

The surface of Mars has in recent years been shown to be  active, and also to exhibit a wide variety of highly localized mineralogical sites of potential interest, for example chloride (salt) minerals in the southern highlands \citep{Osterloo2008}. There are numerous instances of meteoritic impact \citep{Malin2006}, and other signs of activity, ranging from possible land slips to changes in surface coloration which might be produced by near-surface water \citep{Malin2006}, as well as an ongoing debate on the presence of methane venting from the Martian subsurface \citep{Mumma2009, Zahnle2011}. This activity together with the possible protections afforded by evolutionary adaption and residence in the subsurface, as discussed in the Introduction, enhances the probability of finding extant or preserved organisms at the surface. If life is now extinct, but was present in the past and its signatures are preserved in the cold subsurface \citep{Landis2001} then we have an opportunity to remotely sense such material in places where the subsurface has recently be revealed through small impact events. Impacts are of particular interest since they reveal quantities of subsurface material forced to the surface and exposed allowing, for a period of time, access to remote sensing of subsurface material from orbit \citep{Malin2006}.

If  activity relevant to biology is present at or near the surface, then circular polarization spectroscopy represents a means by which signatures of chirality may be remotely sensed. Strong signatures of chirality can arise in the presence of extant life, or from extinct but preserved life if the racemization timescale is lengthy \citep{Landis2001}.
The advantages of a remote sensing spectropolarimetric survey over in situ measurement are primarily in the extent of the surface which may be explored. Above, we discussed the limitations imposed by ground based resolution, and typical dilution tolerances that would be acceptable. From the ground, we can only survey regions of order a few tens of kilometers in extent, at best, however from Mars orbit, resolutions of order tens of {\it meters\/} can be achieved, over an area limited only by the mission duration. Clearly spatial measurements at such a resolution offer a dramatic enhancement of our ability to discriminate chiral signatures, if present, from their surroundings.

A remote sensing survey to seek evidence of the presence of microbes will function most effectively if they are tightly packed and localized. If they are dispersed thinly over a large area, the dilution of any signal will be too great for detection. However, although it is often assumed that the density of any extant microbial organisms on Mars will be very low, in the case of terrestrial bacteria, there is a strong preference for organisms to grow and thrive in close proximity to one another. The majority of terrestrial microbes live in biofilms or similar aggregates  \citep{Costerton1995}. \citet{Shapiro1988} pointed out that our tendency to view bacteria as simple unicellular organisms is likely to be inaccurate, arising from biases of medical bacteriological research, and that in fact bacteria typically function in large ensembles analogous to multicellular organisms. \citet{Costerton1995}, \citet{Davey2000} point out that the majority of microbes persist within a structured biofilm and that biofilms predominate numerically in ecosystems with adequate nutrients. As well as the obvious tendency of replication of a relatively immobile population to produce localized communities, collective activity offers many advantages to the microbial community. These include the ability to proliferate more effectively, to take advantage of localized niches of nutrients, and to build symbiotic relationships with other bacteria utilizing a variety of metabolic processes \citep{Davey2000}. Biofilms are evident very early in the fossil record, appearing over  $3\times 10^9$~years ago and ``biofilm formation is an ancient and integral component of the prokaryotic life cycle, and a key factor for survival'' \citep{HallStoodley2004} .

Hence, we postulate that were microbial life to have evolved on Mars and persisted to survive in the proximity of the surface, it is likely to occur in tight knit communities, to take advantage of localized patches of nutrient, and of the mutual survival strengths that arise in a typical symbiotic environment analogous to a terrestrial microbial biofilm. This would be advantageous for a remote sensing survey of the Mars surface. Quantitatively, in the event that microbial communities are present on scales of tens of meters, and we can resolve such a scale with our spectropolarimeter from Mars orbit, then we require a sensitivity of polarization measurement of $\approx 10^{-4}$ for secure detection of an intrinsic polarization signal of $10^{-3}$.

Space qualified high-sensitivity polarimeters have not yet been developed that could be used for such a mission, though a number are in test. However, a new method of acquiring polarimetric observations with static optics, yielding full Stokes polarimetry on a single data frame, has been advanced by \citet{Sparks2012ao}. By introducing a variation in retardance orthogonal to the spectrum, followed by a polarization analyzer such as a Wollaston prism, it is possible to encode the polarimetric information onto a two dimensional data frame, Fig.~3. In this scheme the polarization Stokes parameters are extracted as the coefficients of orthogonal trigonometric functions which describe the resulting intensity modulation. Fig.~3 shows an implementation of this concept for linear polarimetry; however it is possible to encode the full Stokes parameter set in similar fashion \citep{Sparks2012ao}.

A spectropolarimeter of this type is ideal for surveying the surface of Mars. The large area detector, presumed to be mounted behind a small telescope, ensures a very large number of photons can be acquired quickly. This is essential in precision polarimetry. A typical CCD well depth is $\sim 10^5$~electrons, hence to accomplish polarimetry to a degree of $\sim 10^{-4}$ we require $10^8$~electrons, or  only 1000 CCD pixels. Modern CCDs have many millions of pixels, hence the light gathering issue is greatly diminished. There are other technical advantages in principle to this approach, such as the reduction in CCD flat-fielding uncertainties and absence of moving parts, but the most obvious advantage beyond solving the photon statistical problem is that the measurement is obtained on a single data frame using a single set of optics. Hence, if the target and camera are in relative motion, we can still measure the polarization. In traditional methods such as moving waveplates, the relative motion causes small viewing differences in images that need to be combined, which limits the accuracy of polarimetry. Equally, if a time average is preferred that also may be obtained with no loss of polarization sensitivity in the new approach.

\section{Summary}
\label{summary}
Photosynthetic microbial organisms are of major importance to astrobiology through the evolutionary advantages accrued by photosynthesis and the easy ``observability'' of photosynthesis. We have shown that the (chiral) photosynthetic pigments produce measurable circular polarization signatures. Given potential habitats for microbial life (extant or in the past) and potential strategies for survival that could yield circular polarization signatures (UV protection using pigments, endolithic growth), we suggest that a sensitive polarization survey of selected regions of the Mars surface is warranted. Very large areas could be covered efficiently and a positive signal would  be of significant interest, becoming a high priority target for follow-up investigations. We have described a new approach to polarimetry that may enable such a survey to be carried out.

\vspace{0.5in}
\noindent
{\bf Acknowledgments:} This work was supported by the US National Science Foundation through award numbers EAR 0747394 (F TR),  and MCB 0605301 (F TR), and the STScI JWST Director's Discretionary Research Fund JDF grant number D0101.90152. STScI is operated by the Association for Universities for Research in Astronomy, Inc., under NASA contract NAS5-26555.

\bibliographystyle{model2-names}

\clearpage

\begin{figure}[tbp]
  \centering
  \includegraphics[width=13cm]{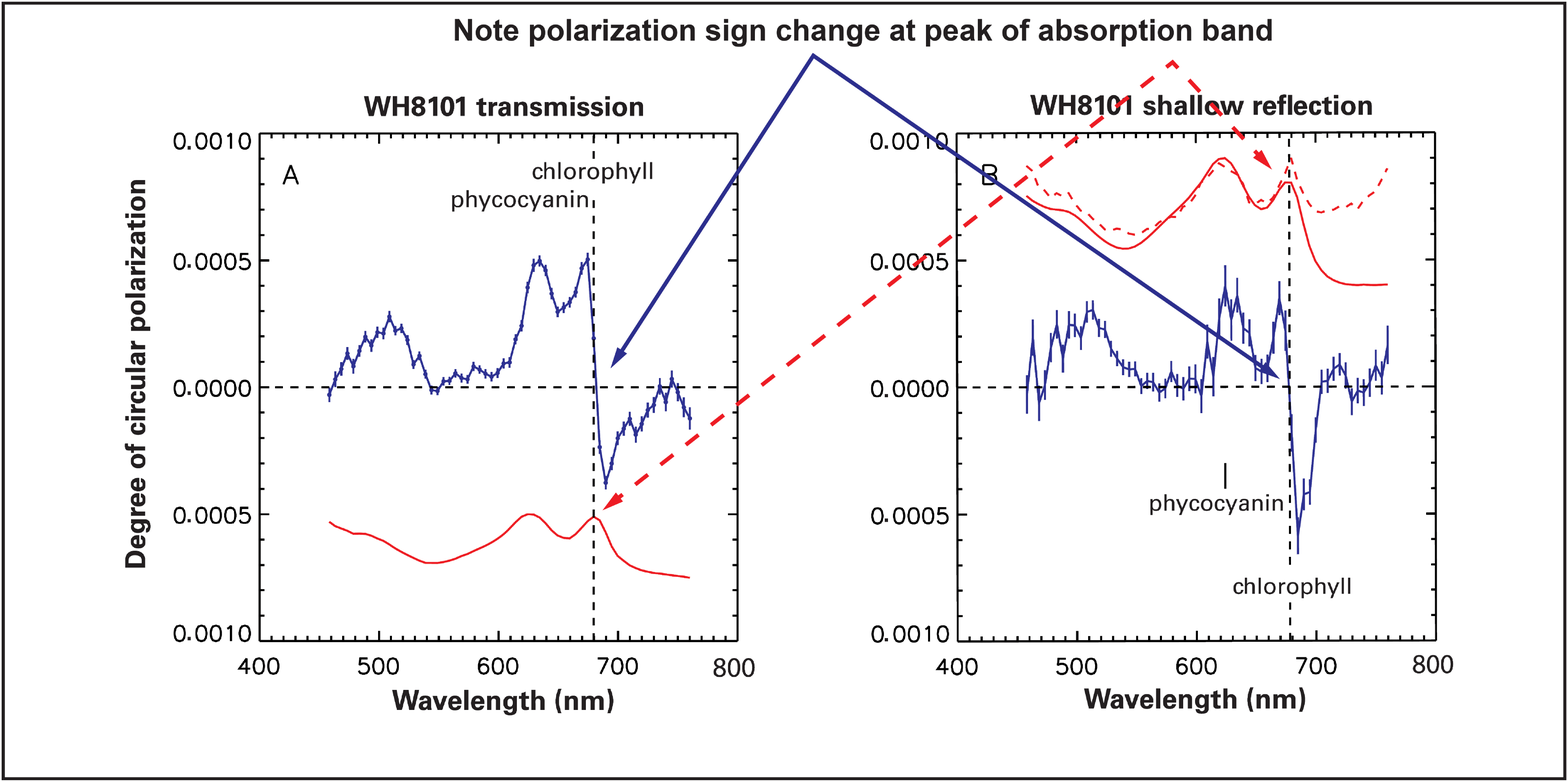}
   \caption{ Photosynthetic cyanobacteria yield a distinctive circular polarization signature (indicated by blue solid arrows) due to the chirality of the photosynthetic molecules. The polarization features correspond to electronic absorption bands (indicated by red dashed arrows) of photosynthesis. The sign-flip at the absorption peak is an extremely distinctive signature of this photosynthetic molecule.}
  \end{figure}

\begin{figure}[htbp]
  \centering
  \includegraphics[width=13.5cm]{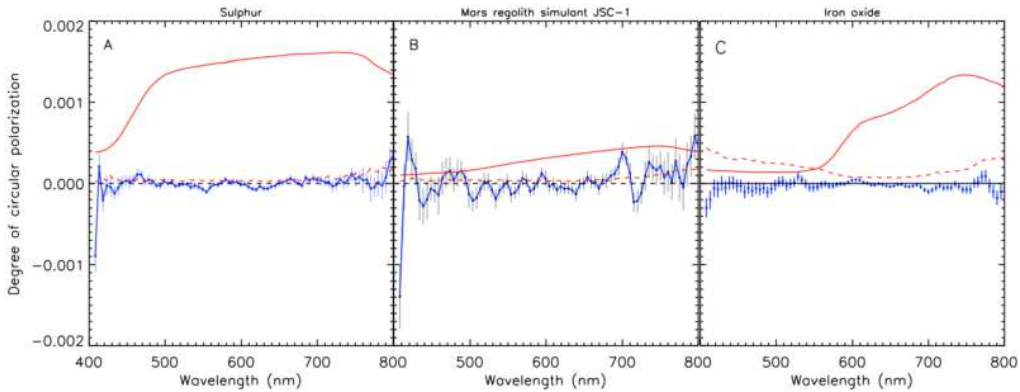}
   \caption{ Minerals sulphur rock, JSC-1 Mars regolith simulant and iron oxide do not show characteristic signatures in circular polarization despite strong spectral signatures. The blue lines with error bars are the circular polarization measurements, the red dashed line is the linear polarization $\times 0.01$ and the solid red lines are  reflectances scaled so that unity, 100\%\ reflectance, is at the top of the graph.}
  \end{figure}

\begin{figure}[htbp]
 \centering
 \includegraphics[width=12cm]{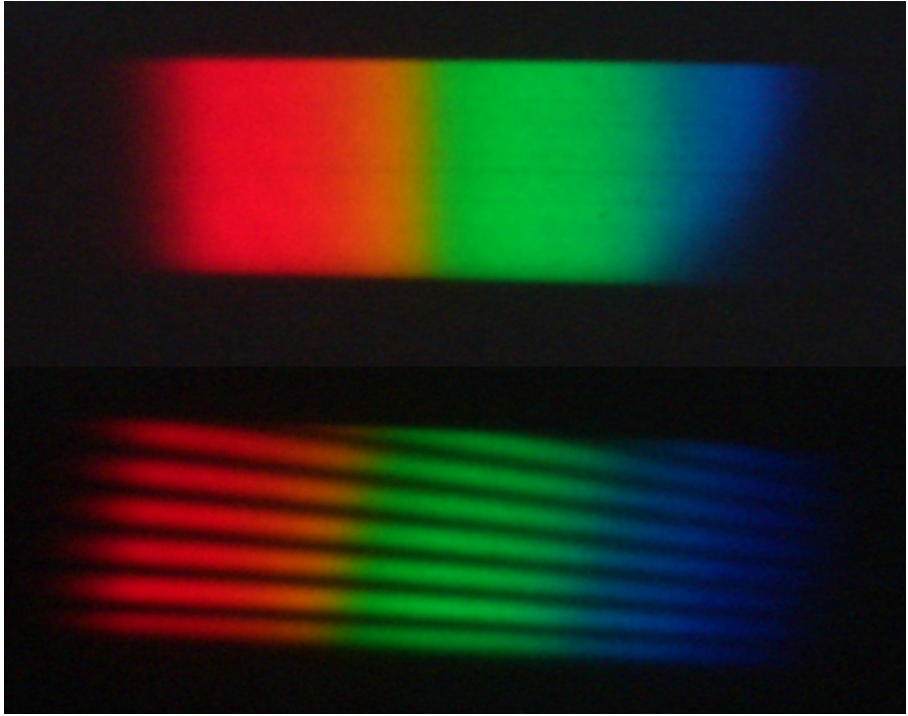}
  \caption{With no moving parts and only static, robust optics, it is possible to encode the polarization information on the spatial dimension of a two dimensional data array. Above shows a two dimensional spectrum with wavelength running horizontally. Below, fringes  appear when polarized light is shone into the spectropolarimeter. Their depth yields the polarization degree and their phase, the direction of polarization. It is possible to encode all Stokes parameters using similar principles \citep{Sparks2012ao}.}
\end{figure}

\end{document}